\documentclass[prd,showpacs,nofootinbib,preprintnumbers,superscriptaddress,twocolumn,prX,superscriptaddress]{revtex4}

\usepackage{graphicx}
\usepackage{amssymb,amsmath}
\usepackage{enumerate}
\usepackage{multirow}

\newcommand{\E}[1]{\times 10^{#1}}
\newcommand{\dee}{\mathrm{d}}

\begin{document}

\title{Determining scalar field potential in power-law cosmology
with observational data}
\author{Kiattisak Thepsuriya}
\email{kiattisak@physics.org}
\affiliation{NEP, The Institute for Fundamental Study, Naresuan University, Phitsanulok 65000, Thailand}

\author{Burin Gumjudpai\footnote{Corresponding author}}
\email{buring@nu.ac.th} \affiliation{Fundamental Physics \& Cosmology Research Unit, The Tah Poe Academia Institute (TPTP), Department of
Physics, Naresuan University, Phitsanulok 65000, Thailand}
\affiliation{Thailand Center of Excellence in Physics, Ministry of Education, Bangkok 10400, Thailand}
\affiliation{NEP, The Institute for Fundamental Study, Naresuan University, Phitsanulok 65000, Thailand}

\begin{abstract}
In power-law cosmology, we determine potential function of a canonical scalar field in FLRW universe in presence of barotropic perfect fluid.
The combined WMAP5+BAO+SN dataset and WMAP5 dataset are used here to determine the value of the potential. The datasets suggest slightly closed
universe. If the universe is closed, the exponents of the power-law cosmology are $q = 1.01$ (WMAP5 dataset) and $q=0.985$ (combined dataset).
The lower limits of $a_0$ (closed geometry) are $5.1\E{26}$ for WMAP5 dataset and $9.85\E{26}$ for the combined dataset. The domination of the
power-law term over the curvature and barotropic density terms is characterised by the inflection of the potential curve. This happens when the
universe is 5.3 Gyr old for both datasets.
\end{abstract}

\pacs{98.80.Cq}

\date{\today}
 \maketitle

\section{Introduction}

The presence of a scalar field is motivated by many ideas in high energy physics and quantum gravities, although it has not been discovered
experimentally. TeV-scale experiments at LHC and Tevatron may be able to confirm its existence. It is nevertheless widely accepted in
several theoretical modeling frameworks, especially in contemporary cosmology, in which an early-time accelerated expansion, i.e., inflation, is
proposed to be driven by a scalar field in order to solve horizon and flatness problems \cite{inflation}. After inflation, components of
barotropic fluids such as radiation and other non-relativistic matter were produced during reheating and cooling-down processes. A scalar field
was also believed to be responsible for the present acceleration in various models of dark energy \cite{Padmanabhan:2004av}. The present
acceleration is strongly backed up by various observations, e.g. the cosmic microwave background \cite{Masi:2002hp}, large-scale structure surveys
\cite{Scranton:2003in} and SNe type Ia observations \cite{Riess:1998cb, Riess:2004nr, Astier:2005qq}.

Power-law cosmology, where $a \propto t^q$, describes an acceleration phase if $q > 1$. Modelling the present expansion with a power-law
function where $q \sim 1$, although tightly constrained by nucleosynthesis \cite{Sethi:1999sq, Kaplinghat:1999}, considering at later time, studies of  age of high-redshift objects
such as globular clusters \cite{Kaplinghat:1999, Lohiya:1997ti,Dev:2002sz, Sethi2005}, SNe Ia data \cite{Sethi2005, Kumar2011}, SNe Ia with $H(z)$ data
\cite{Dev2001, Sethi2005, Dev:2008ey} and X-ray gas mass fraction measurement of galaxy clusters \cite{Allen2002, Zhu:2007tm} in context of power-law cosmology are well-viable. Moreover other aspects such as gravitational lensing statistics \cite{Dev:2002sz}, angular size-redshift data of compact radio sources
\cite{Jain2003} have also been studied in power-law cosmology. Originally, the power-law expansion has
its motivation from the simplest inflationary model that can remove the flatness and horizon problems with simple spectrum \cite{Lucchin}. For
the present universe, the idea of linear coasting cosmology ($a \propto t$) \cite{Kolb1989} can resolve the age problem of the CDM model
\cite{Lohiya:1997ti} while as well agreeing with the nucleosynthesis constraint. The coasting model arises from non-minimally coupled
scalar-tensor theory in which the scalar field couples to the curvature to contribute to the energy density that cancels out the vacuum energy
\cite{Sethi:1999sq, Ford1987}. The model could also be a result of the domination of an SU(2) cosmological instanton \cite{Allen1999}.

Here our assumption is that the universe is expanding in the form of the power law function. Two major ingredients are scalar field dark energy
evolving under the scalar field potential $V(\phi)$, and barotropic fluid consisting of cold dark matter and baryons. We derive the potential,
and use the combined WMAP5 data \cite{Hinshaw:2008kr} as well as the WMAP5 data alone to determine the values of $q$ and other relevant
parameters of the potential. The numerical results are subsequently compared and discussed.

\section{Cosmological System with Power-Law Expansion}

Two perfect fluids, the cold dark matter and scalar field $\phi \equiv \phi(t)$, in the late FLRW universe of the simplest CDM model with zero
cosmological constant are considered. The time evolution of the barotropic fluid is governed by the fluid equation
\begin{equation}
\dot{\rho}_\gamma =
-3H\rho_\gamma,
\end{equation}
 since  $w_{\gamma}$ is constant,
\begin{equation}\label{energydensity} \rho_\gamma = \frac{D}{a^n}, \end{equation}
where $n \equiv 3 (1 + w_\gamma)$ and $D \geq 0$ is a proportional constant. For the scalar field, supposed that it is minimally coupled to
gravity, its Lagrangian density is $\mathcal{L} =  \dot{\phi}^2 /2 - V(\phi)$. The energy density and pressure are
\begin{align}
\rho_\phi = \frac{1}{2}  \dot{\phi}^2 + V(\phi),\quad  p_\phi = \frac{1}{2}  \dot{\phi}^2 - V(\phi).
\end{align}
The fluid equation of the field describing its energy conservation as the universe expands is
\begin{equation}  \ddot{\phi} + 3H\dot{\phi} + \frac{\dee}{\dee \phi}V = 0. \end{equation}
Total energy density $\rho_\mathrm{tot}$ and total pressure $p_\mathrm{tot}$ of the mixture are simply the sums of those contributed by each
fluid, for which the Friedmann equation is
\begin{equation}
H^2 = \frac{8 \pi G}{3} \rho_\mathrm{tot} - \frac{k c^2}{a^2}.
\end{equation}
It is straightforward to show that
\begin{equation}\label{vraw} V(\phi) = \frac{3}{8 \pi G} \left( H^2 + \frac{\dot{H}}{3} + \frac{2k}{3a^2} \right) + \left( \frac{n - 6}{6} \right) \frac{D}{a^n}, \end{equation}
where $8\pi G$ is related to the reduced Planck mass $M_\mathrm{P}$ by $8\pi G = M_\mathrm{P}^{-2}$. The power-law scale factor is
\begin{equation}\label{scalefactor} a(t) = a_0 \left( \frac{t}{t_0} \right)^q\,, \end{equation}
without fixing $a_0 = 1$ at the present time because we have implicitly rescaled it to allow for $k$ taking only either one of the three
discrete values $0, \pm1$. The Hubble parameter is
\begin{equation}\label{hubbleparameter} H(t) = \frac{\dot{a}(t)}{a(t)} = \frac{q}{t}. \end{equation}
Our goal is to construct $V(t)$ using recent observational data, as far as the simplest CDM model is concerned.

\section{Scalar Field Potential}
We will work with observational data in SI units. Restoring the physical constants in place, we obtain
\begin{equation} V(\phi) = \frac{3 M_\mathrm{P}^2 c}{\hbar} \left( H^2 + \frac{\dot{H}}{3} + \frac{2kc^2}{3a^2} \right) - \frac{Dc^2}{2a^3}, \end{equation}
where $M_\mathrm{P}^2 =\hbar c / 8\pi G$ and we have set $n = 3$ ($w_\gamma = 0$ for dust). Incorporating \eqref{scalefactor} and \eqref{hubbleparameter} into the above equation, we obtain
\begin{equation}
 V(t) = \frac{M_\mathrm{P}^2 c}{\hbar}
 \left( \frac{3q^2 - q}{t^2} + \frac{2kc^2 t_0^{2q}}{a_0^2 t^{2q}} \right) - \frac{Dc^2}{2}\frac{t_0^{3q}}{a_0^3 t^{3q}}. \label{vt}
 \end{equation}
We shall consider contribution of the first term alone in comparison to total contribution when including the second (the curvature) and the
third (density) terms.  It is worth noting that there are previous attempts using theoretical methods to construct scalar field potential in various cases, e.g. assuming scaling solution \cite{astro-ph/9809272}, case of negative potential of scale field without using observational data \cite{Cardenas:2004ji} and construction of scalar field potential using dark energy density function with non-specified expansion law \cite{Guo:2006ab}. Constructions of the potential using  SNe Ia dataset were studied \cite{Li:2006ea}.

\subsection{Cosmological Parameters}

Using the equation for the Hubble parameter \eqref{hubbleparameter} at the present time, we have
\begin{equation} q = H_0 t_0. \end{equation}
The sign of $k$ depends on the sign of the density parameter $\Omega_k \equiv -kc^2 / a^2 H^2$. In our convention here, $k = 1 \;(\Omega_{k} <
0)$ for a closed universe, $k = 0$ for a flat one, and $k = -1 \;(\Omega_{k} > 0)$ for an open one. The present value of the scale factor can be
found from the definition of $\Omega_{k,0}$, that is,
\begin{equation} a_0 = \frac{c}{H_0} \sqrt{\frac{-k}{\Omega_{k, 0}}}. \end{equation}
The density constant $D$ can be found from  \eqref{energydensity},
\begin{equation} D = \rho_{\gamma, 0} a_0^3 = \Omega_{\gamma, 0} \rho_{c, 0} a_0^3, \end{equation}
where $ \Omega_{\gamma, 0} = \Omega_{\mathrm{CDM}, 0} + \Omega_{b, 0},$ i.e. the sum of the present density parameters of the barotropic fluid
components. $\rho_{c, 0}$ is the present value of the  critical density. The neutrino contribution is assumed to be negligible. The values of
$H_0$, $t_0$, $\Omega_{k, 0}, \Omega_{\mathrm{CDM}, 0}$, and $\Omega_{b, 0}$ are taken from observational data.

\subsection{Observational Data}

We work on two sets of data provided by \cite{Hinshaw:2008kr}. One comes solely from the WMAP5 data and the other is the WMAP5 data combined
with distance measurements from Type Ia supernovae (SN) and the Baryon Acoustic Oscillations (BAO) in the distribution of galaxies. For $t_0$,
$H_0$, $\Omega_{b, 0}$, and $\Omega_{\mathrm{CDM}, 0}$, we take their maximum likelihood values. The curvature density parameter $\Omega_{k, 0}$
comes as a range with 95\% confidence level on deviation from the simplest $\Lambda$CDM model. The data are shown in Table \ref{datatable}.

\begin{table*}[t]
\begin{ruledtabular}
\begin{tabular}{ccc}
\textbf{Parameter} & \textbf{WMAP5+BAO+SN} & \textbf{WMAP5}\\
\hline
$t_0$ & $13.72$ Gyr & $13.69$ Gyr\\
$H_0$ & $70.2$ km/s/Mpc & $72.4$ km/s/Mpc\\
$\Omega_{b, 0}$ & $0.0459$ & $0.0432$\\
$\Omega_{\mathrm{CDM}, 0}$ & $0.231$ & $0.206$\\
$\Omega_{k, 0}$ & $-0.0179 < \Omega_{k, 0} < 0.0081$ & $-0.063 < \Omega_{k, 0} < 0.017$\\
\end{tabular}
\caption{Observational data used in the construction of our scalar-field potentials \cite{Hinshaw:2008kr}} \label{datatable}
\end{ruledtabular}
\end{table*}

\section{Results and Discussions}

\begin{figure}
\centering
\includegraphics[width=3.2in]{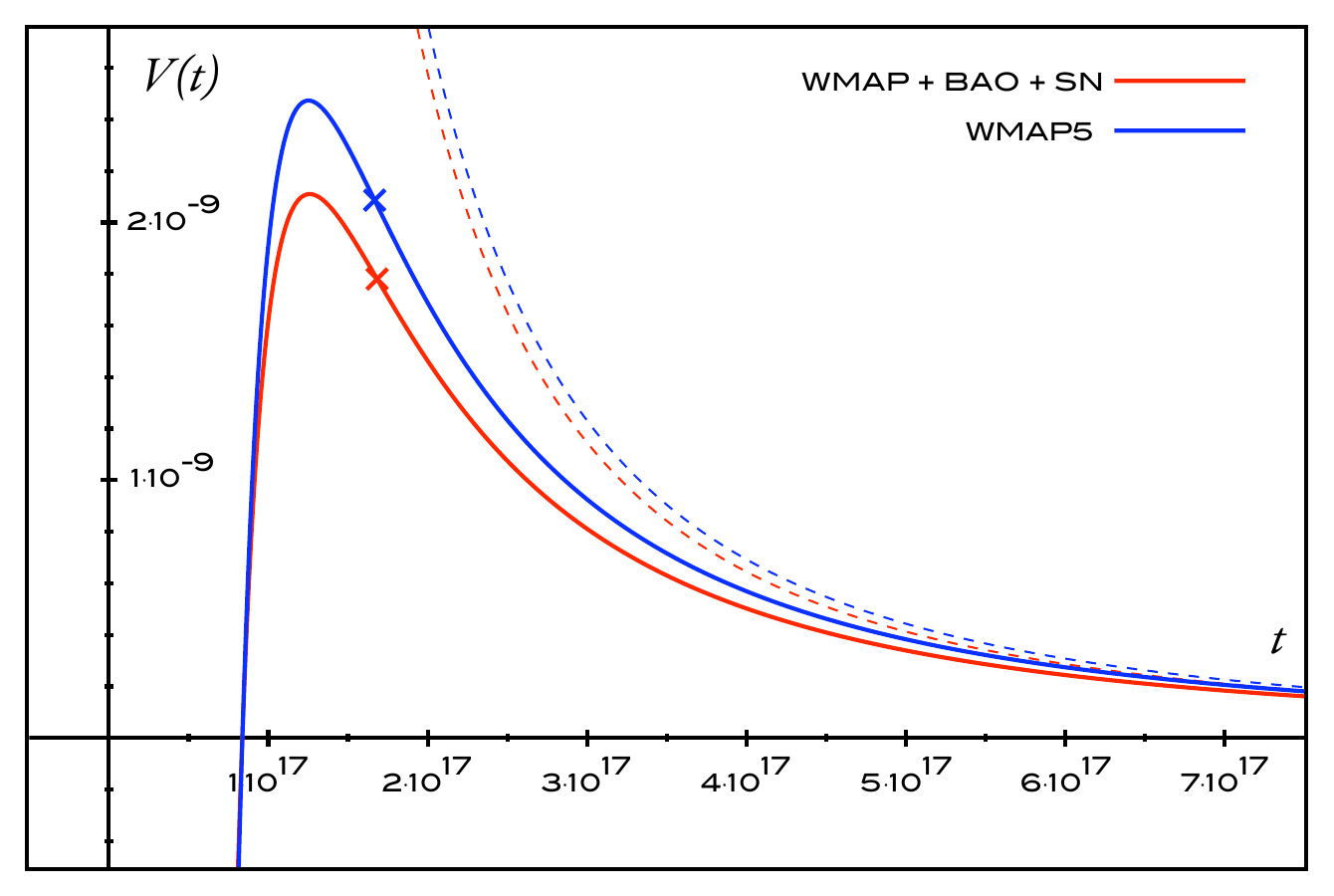}
\caption{The potentials in \eqref{firstpotential} and \eqref{secondpotential}. The units of the abscissa and ordinate axes are sec and
J$/$m${}^3$, respectively. The crosses mark their inflection points. Also plotted in dash lines are their first terms. Each potential does not
actually converge to its first term, but later intersect with and deviate from it, though still very close together. However, this occurs much
later (at $t = 2.8\E{84}$ sec $= 8.8\E{67}$ Gyr in both cases). } \label{potentialplots}
\end{figure}

Using combined WMAP5+BAO+SN dataset, the potential is
\begin{equation}\label{firstpotential} V(t) = \frac{1.03\E{26}}{t^2} + \frac{1.5\E{23}}{t^{1.97}} - \frac{1.5\E{42}}{t^{2.96}}, \end{equation}
whereas, for WMAP5 dataset alone,
\begin{equation}\label{secondpotential} V(t) = \frac{1.11\E{26}}{t^2} + \frac{7.6\E{24}}{t^{2.03}} - \frac{4.6\E{43}}{t^{3.04}}. \end{equation}
in SI units. We use the mean of each $\Omega_{k, 0}$ interval to represent $\Omega_{k, 0}$ in each of the above equations. Their plots are shown
in Fig. \ref{potentialplots}. In both cases, $\bar{\Omega}_{k, 0}$ is negative (a closed universe). The points at which the potential, its
derivative, and its second-order derivative, are zero ($t_\textrm{intercept}$, $t_\textrm{max}$, and $t_\textrm{inflection}$, respectively) are
also determined, for both $\bar{\Omega}_{k, 0}$ and each end of the $\Omega_{k, 0}$ interval. The results are summarised in Table
\ref{resulttable}.

\begin{table*}[t]
\begin{ruledtabular}
\begin{tabular}{ccccc}
& \multicolumn{2}{c}{\textbf{WMAP5+BAO+SN}} & \multicolumn{2}{c}{\textbf{WMAP5}}\\
& $\bar{\Omega}_{k, 0} = -0.0049$ & $-0.0179 < \Omega_{k, 0} < 0.0081$ & $\bar{\Omega}_{k, 0} = -0.023$ & $-0.063 < \Omega_{k, 0} < 0.017$\\
\hline
$q$ & $0.985$ & $0.985$ & $1.01$ & $1.01$\\
\multirow{2}{*}{$a_0$} & \multirow{2}{*}{$1.9\E{27}$} & $a_0 > 9.85\E{26}$ (closed) & \multirow{2}{*}{$8.4\E{26}$} & $a_0 >5.1\E{26}$ (closed)\\
&& $a_0 > 1.5\E{27}$ (open) && $a_0 > 9.8\E{26}$ (open)\\
$t_\textrm{intercept}$ & $2.7$ Gyr & $2.62$ Gyr $< t < 2.7$ Gyr & $2.7$ Gyr & $2.6$ Gyr $< t < 2.8$ Gyr\\
$t_\textrm{max}$ & $4.0$ Gyr & $3.94$ Gyr $< t < 4.0$ Gyr & $4.0$ Gyr & $3.8$ Gyr $< t < 4.1$ Gyr\\
$t_\textrm{inflection}$ & $5.3$ Gyr & $5.26$ Gyr $< t < 5.4$ Gyr & $5.3$ Gyr & $5.1$ Gyr $< t < 5.5$ Gyr\\
\end{tabular}
\caption{A summary of numerical results. Times are shown in Gyr for comprehensibility. Positive and negative $\Omega_k$'s correspond to open and
closed universes, respectively.} \label{resulttable}
\end{ruledtabular}
\end{table*}
The values of the exponent $q$ from the two sets of data are only slightly different, but only the latter is an accelerated expansion as $q > 1$.
 The determination of $q$ from X-Ray gas mass fractions
in galaxy clusters favours open universe with $q > 1$ ($q = 1.14 \pm 0.05$) \cite{Zhu:2007tm}  and combined analysis from SNLS and $H(z)$ data
(from Germini Deep Deep Survey) assuming open
 geometry yields $q = 1.31$ \cite{Dev:2008ey}. Note that, in the power-law regime, $q$ only depends on the observed values of the Hubble constant and
$t_0$. This may give an impression that the maximum likelihood values from the combined data has yet to be relied upon, but the power-law
expansion has not been proven to be the case nonetheless.

\begin{figure}
\centering
\includegraphics[width=3.2in]{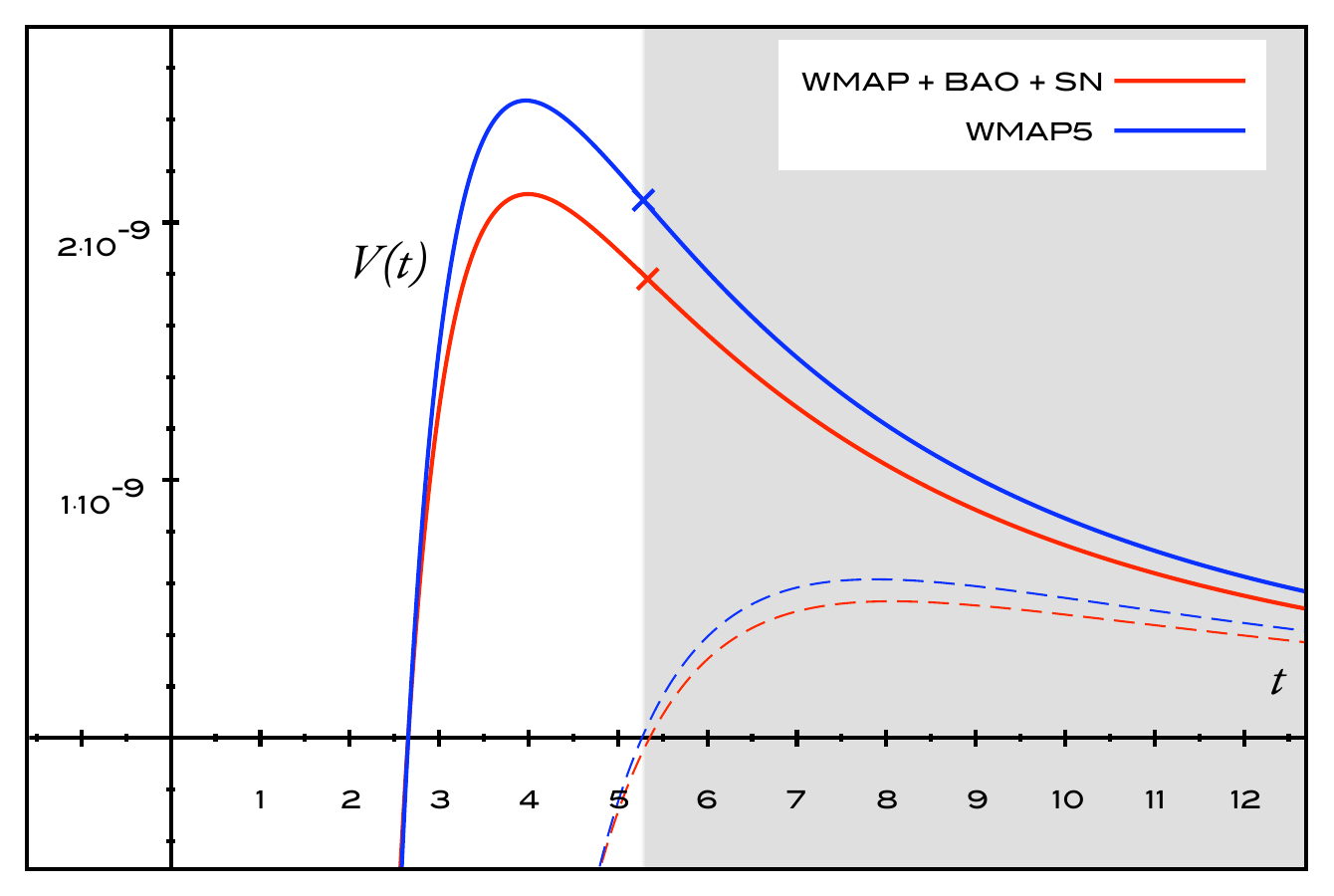}
\caption{The potentials in \eqref{firstpotential} and \eqref{secondpotential} along with the radicands of the integrands in
\eqref{firstscalarfield} and \eqref{secondscalarfield} (dash line). The shaded region is the post-inflection phase. The unit of the abscissa
axis is sec. After $t_\textrm{inflection}$, $\phi(t)$ is real.} \label{scalarfieldplots}
\end{figure}
After $t_\textrm{inflection}$, the potential from each data behaves like its first term, i.e. decreasing in its value while increasing in its
slope (being less and less negative). The other terms quickly become weaker. This can be seen in Fig. \ref{potentialplots}. Since the first term
is contributed only by $H(t)$ (and its time derivative), it is dominant in the post-inflection phase. In fact, the convergence to zero of the
potential is slower than its first term alone (see (\ref{firstpotential}) and (\ref{secondpotential})), because the sum of the last two terms
consequently becomes positive before converging to zero. This means that the plots of each potential and its first term in Fig.
\ref{potentialplots} eventually crosses, but it occurs much, much later at $t = 8.8\E{67}$ Gyr. Along with the potential function in
\eqref{vraw}, we also obtain the solution
\begin{equation} \phi(t) = \int \sqrt{- \frac{2 M_\mathrm{P}^2 c}{\hbar} \left( \dot{H} - \frac{kc^2}{a^2} \right) - \frac{Dc^2}{a^3}} \, \dee t \end{equation}
in SI units. Using WMAP5+BAO+SN dataset,
\begin{equation}\label{firstscalarfield} \phi(t) = \int \sqrt{\frac{1.06\E{26}}{t^2} + \frac{1.5\E{23}}{t^{1.97}} - \frac{3.0\E{42}}{t^{2.96}}} \, \dee t, \end{equation}
where, for WMAP5 dataset alone,
\begin{equation}\label{secondscalarfield} \phi(t) = \int \sqrt{\frac{1.09\E{26}}{t^2} + \frac{7.6\E{24}}{t^{2.03}} - \frac{9.3\E{43}}{t^{3.04}}} \, \dee t. \end{equation}
In the late post-inflection phase, the first term is dominant over the $k$ and $D$ terms then the last two terms of the radicands are negligible
(Fig. \ref{scalarfieldplots}). The above two equations are approximated as
\begin{equation} \phi(t) \approx 1.04\E{13} \ln t, \end{equation}
whereas, for WMAP5 dataset alone,
\begin{equation} \phi(t) \approx 1.03\E{13} \ln t. \end{equation}
The radicand in (\ref{secondscalarfield}) of the WMAP5 dataset is zero at approximately $t_\textrm{inflection} = 5.3$ Gyr (see Fig.
\ref{scalarfieldplots}), therefore so does $\phi(t)$. While the combined dataset has the zero radicand (then zero $\phi(t)$) in
(\ref{firstscalarfield}) later at approximately $t = 5.4$ Gyr. Scalar field exact solutions for the power-law cosmology with non-zero curvature
and non-zero matter density are reported in \cite{NLS2}. It is also worth noting that the general exact form of the potential, that renders
scaling solution, is some negative powers of a hyperbolic sine \cite{Rub:2001}.
\vspace{0.2cm}

\section{Conclusion}
We consider a potential function of a homogeneous scalar field in late-time FLRW universe of the simplest CDM model with zero cosmological
constant, assuming power-law expansion. The scalar field is minimally coupled to gravity and the other fluid is non-relativistic barotropic
perfect fluid. We use two sets of observational data, combined WMAP5+BAO+SN dataset and WMAP5 dataset, as the inputs. Potential functions are
obtained using numerical values from the observations. Mean values of both sets suggest slightly closed geometry. The WMAP5 dataset implies
accelerated expansion ($q = 1.01$) while the combined dataset gives $q=0.985$. This is slightly lower than the value obtained from SNLS and
$H(z)$ data ($q = 1.31$) \cite{Dev:2008ey}  and X-Ray gas mass fraction ($q=1.14\pm 0.05$) \cite{Zhu:2007tm}. Our result is independent of the
geometry unlike $q$ obtained from \cite{Dev:2008ey} which assumes open geometry.
 For closed universe, the WMAP5
dataset puts the lower limit of $5.1\E{26}$ for $a_0$ while the combined dataset puts the lower limit of $9.85\E{26}$. We characterise the
domination of the first term of (\ref{vt}) by using the inflection of the potential plots from which the first term is found to be dominant to
the potential 5.3 Gyr after the Big Bang in both datasets.

\section*{Acknowledgments}
We thank Chris Clarkson for discussion. K.~T. is supported by a research assistantship under a grant of the Thailand Toray Science
Foundation (TTSF) and a NARIT postgraduate studentship. B.~G. is sponsored by the Thailand Research Fund and the National Research Council of Thailand.

\end{document}